\documentclass[a4paper,USenglish]{lipics}

\usepackage{microtype}

\usepackage{listings}
\usepackage[colorinlistoftodos]{todonotes}
\usepackage{booktabs}

\usepackage{paralist}
\usepackage{array}
\usepackage{collcell}

\usepackage{tikz}
\usetikzlibrary{shapes,arrows,fit,backgrounds,calc,positioning}

\lstdefinelanguage{JavaScript}{
  keywords={attributes, class, classend, do, empty, endif, endwhile, fail,
  function, functionend, if, implements, in, inherit, inout, not, of,
  operations, out, return, then, types, while, use, else, switch, case,
  break, default, for, var, new, throw, void, yield, typeof, try,
  this, instanceof},
  keywordstyle=\color{black}\bfseries,
  ndkeywords={},
  ndkeywordstyle=\color{black}\bfseries,
  identifierstyle=\color{black},
  sensitive=false,
  comment=[l]{//},
  morecomment = [s]{/*}{*/},
  morecomment = [s][\color{gray}]{/**}{*/},
  commentstyle=\color{gray},
  stringstyle=\color{black}
}
\newcolumntype{I}{>{\iffalse}c<{\fi}@{~}}

\newcolumntype{H}{>{\setbox0=\hbox\bgroup}c<{\egroup}@{}}

\usepackage{pifont}
\newcommand{\typeI}{\textbf{Type-I}}
\newcommand{\typeIa}{\textbf{Type-Ia}}
\newcommand{\typeIb}{\textbf{Type-Ib}}
\newcommand{\typeII}{\textbf{Type-II}}
\newcommand{\typeIIa}{\textbf{Type-IIa}}
\newcommand{\typeIIb}{\textbf{Type-IIb}}

\title{Transparent Object Proxies for JavaScript%
  \newline\textnormal{Technical Report}
}
\titlerunning{Transparent Object Proxies for JavaScript}

\author[1]{Matthias Keil}
\author[2]{Sankha Narayan Guria}
\author[1]{Andreas Schlegel}
\author[1]{Manuel Geffken}
\author[1]{Peter Thiemann}

\affil[1]{Institute for Computer Science\\University of Freiburg\\Freiburg, Germany\\
  \texttt{\{keilr,schlegea,geffken,thiemann\}@informatik.uni-freiburg.de}
}
\affil[2]{Indian Institute of Technology Jodhpur\\Jodhpur, India\\
  \texttt{sankha@iitj.ac.in}
}

\authorrunning{M. Keil, S. N. Guria, A. Schlegel, M. Geffken, and P. Thiemann} 

\Copyright{Matthias Keil, Sankha Narayan Guria, Andreas Schlegel, Manuel Geffken, and Peter Thiemann}



\begin{document}

\maketitle

\begin{abstract}
  Proxies are the swiss army knives of object adaptation. They
  introduce a level of indirection to intercept select operations on
  a target object and divert them as method calls to a handler.
  Proxies have many uses like implementing access control, enforcing
  contracts, virtualizing resources.

  One important question in the design of a proxy API is whether a
  proxy object should inherit the identity of its target. Apparently
  proxies should have their own identity for security-related
  applications whereas other applications, in particular contract
  systems, require transparent proxies that compare equal to their
  target objects.

  We examine the issue with transparency in various use cases for proxies, discuss different approaches to
  obtain transparency, and propose two designs that require modest
  modifications in the JavaScript engine and cannot be bypassed
  by the programmer.

  We implement our designs in the SpiderMonkey
  JavaScript interpreter and bytecode compiler. 
  Our evaluation shows that these modifications of have no statistically
  significant impact on the benchmark performance of the JavaScript
  engine. Furthermore, we demonstrate that contract systems based on
  wrappers require transparent proxies to avoid interference with program execution in
  realistic settings. 

  

  


\end{abstract}



\section{Introduction}
\label{sec:introduction}

A proxy modifies the functionality of an underlying target object by introducing a level of indirection that intercepts all operations on the target. 
As all problems in computer science can be solved by another level of
indirection\footnote{A famous quote by David Wheeler.}, proxies may be called
the Swiss army knives of object adaptation.
Indeed, proxies are widely used to perform resource management, to access remote
objects, to impose access
control~\cite{CutsemMiller2010,KeilThiemann2013-Proxy},
to implement contract checking~\cite{StricklandTobinHochstadtFindlerFlatt2012,FindlerFelleisen2002,Disney2013:contracts},
to restrict the functionality of an object~\cite{StricklandTobinHochstadtFindlerFlatt2012}, 
to enhance the interface of an object~\cite{WarthStanojevicMillstein2006},
to implement dynamic effect systems, 
as well as for meta-level extension,
behavioral reflection, security, 
and concurrency control~\cite{MillerCutsemTulloh2013,AustinDisneyFlanagan2011,BrachaUngar2004}.

A proxy implementation
provides an intercession API that enables the programmer to trap all
operations on the target object (with few exceptions).
Further, a program should not be able to distinguish a proxy from
a non-proxy object so that
putting a proxy in place of an object does
not affect the outcome of the program (save for the new behavior
introduced by the proxy). For that reason, the JavaScript Proxy API
\cite{CutsemMiller2010}, a part of the current ECMAScript 6 draft standard,  does not include a
function that checks whether an object is a proxy, it does not
provide traps for all operations on objects, and it restricts some
traps to avoid breaking certain object invariants
\cite{CutsemMiller2013}.

However, the JavaScript Proxy API embodies a design decision that 
reveals the presence of proxies in some important use
cases. This decision concerns object equality. The API
description\footnote{%
  \scriptsize\url{https://developer.mozilla.org/en/docs/Web/JavaScript/Reference/Global_Objects/Proxy}
} says: \emph{%
  The double and triple equal (\lstinline{==},
  \lstinline{===}) operator is not trapped. \lstinline{p1 === p2} if
  and only if \lstinline{p1} and \lstinline{p2} refer to the same proxy.
} 
The standard does not even mention proxies in the definition of object
equality: \emph{If x and y are the same Object value, return true.}  \cite[Section 7.2.13]{ecmascript2015}
In other words, proxies are \emph{opaque}, which
means that each proxy has its own identity, different from all other
(proxy or non-proxy) objects. Given opaque proxies, an equality test
can be used to distinguish a proxy from its 
target as demonstrated in Listing~\ref{lst:proxy}.
\begin{lstlisting}[name=proxy,float,label={lst:proxy},caption={Comparing a proxy with
  its target. The methods of \lstinline{handler}
  determine the behavior of \lstinline{proxy}. If 
  \lstinline{handler} is empty, then \lstinline{proxy} behaves
exactly like \lstinline{target}.}]
var target = { /* some object */ };
var handler = { /* empty handler */ };
var proxy = new |Proxy| (target, handler);
proxy===target; // evaluates to false
\end{lstlisting}
Even though \lstinline{target} and \lstinline{proxy} behave identically, 
they are not considered equal. Thus, in a program that uses object
equality, the introduction of a proxy along one execution
path may change the meaning of the
program without even invoking an operation on the proxy (which may
behave differently from the same operation on the target).

Equality for opaque proxies is straightforward to implement and works well under the assumption that proxies and
their targets are never part of the same execution
environment. For example, the revocable membrane pattern \cite{CutsemMiller2010} enables to
safely pass object references to untrusted code, control their
operation on these objects, and revoke all access rights afterwards.
This pattern is implemented using proxies and it partitions the
execution environment into security realms so that the objects that
live in the same realm are never in a proxy-target relationship. By
this convention, the situation outlined in Listing~\ref{lst:proxy}
never arises inside a compartment. 



But the assumption that proxies never share their execution
environment with their targets is not always appropriate. One prominent use case
is the implementation of a contract system. A contract system provides
a domain specific language to state very precise type-like assertions for
values in an untyped language. Two examples for such systems are
the contract framework of Racket \cite[Chapter 7]{FlattFindlerPlt2014:_racket_guide} and
Contracts.js for JavaScript \cite{Disney2013:contracts}. Both systems implement
contracts on objects with specific wrapper objects, Racket's chaperones or impersonators
\cite{StricklandTobinHochstadtFindlerFlatt2012} and JavaScript
proxies, respectively. 

\begin{lstlisting}[name=proxy,float=t,label={lst:simple-object-contract},caption={Application with contract wrapper}]
// contract wrapper implementation
function checkPredicate (pred) {
  return {
    set: function(target, prop, val) {
      if (!pred (val)) { throw new ContractException(); };
      target[prop] = val;
    } 
  }
};
function assertContract(target, pred) {
  return new |Proxy| (target, checkPredicate(pred));
}
// application code
function addBonus (acc1, acc2, amount) {
  acc1.balance += amount;
  if (acc1 !== acc2) { // test objects for equality*'\label{line:transfer-equality}'*
    acc2.balance += amount;
  }
}
var account = { balance: 10 };
var restricted = assertContract (account, function(x) {
  return (x >= 0);
});
addBonus (account, account, 40);  // raises account by 40*'\label{line:transfer-ok}'*
addBonus (restricted, account, 40);  // raises account by 80*'\label{line:transfer-broken}'*
\end{lstlisting}
Listing~\ref{lst:simple-object-contract} contains a JavaScript
implementation of a simple contract
wrapper. The implementation uses JavaScript proxies,
which are introduced in more detail in Section~\ref{sec:proxies}. The
wrapper applies to a JavaScript object and it enforces that all property values
written to the object fulfill a predicate \lstinline{pred}.
Otherwise, the wrapper raises an exception.

The call \lstinline{assertContract (target, pred)}
returns a proxy for the target object with a handler created by the
function call \lstinline{checkPredicate (pred)}. 
Whenever a property \lstinline{prop} is set on the proxy,
the method \lstinline{set} of its handler is invoked with the target 
object, the property \lstinline{prop}, and the new value as
arguments. The handler throws an exception if the predicate \lstinline{pred} is not
fulfilled, otherwise it performs the set operation on the target.

The application code contains an \lstinline{addBonus} function that
takes two accounts and a bonus amount to add. The intention is to give
a bonus to each account once. Thus, if the two accounts are
different, then the balance of the second account must be adjusted, too.

The last couple of lines create an account and a restricted handle to
the same account, where the restricted handle does not permit the
account to overdraw. In line~\ref{line:transfer-ok}, a
bonus of \lstinline{40} is added to \lstinline{account} and \lstinline{account}. This
bonus addition executes correctly because the equality test in
line~\ref{line:transfer-equality} yields \lstinline{false}. However, performing the
bonus addition with the restricted version and the standard account
leads to adding a bonus of \lstinline{80} to \lstinline{account} because the
test in line~\ref{line:transfer-equality} yields \lstinline{true}.

This example shows that the introduction of a contract monitor like
\lstinline{assertContract} may change the semantics of a program even
in cases where the contract 
is not exercised. But this change in behavior violates a ground rule for monitoring: a
monitor should never interfere with a program conforming to the
monitored property. (In Section~\ref{sec:usecases}, we make a similar
case for access restricting membranes.)

While this particular example is constructed we demonstrate in
Section~\ref{sec:evaluation} that such situations do occur in
practice. Racket programmers
have also run into this issue\footnote{Personal communication with
Robby Findler, February 2015.}, as chaperones and impersonators 
behave opaquely with respect to Racket's \texttt{eq?} operation.

As a remedy, we propose alternative designs for \emph{transparent
proxies} that are better suited for use cases such as certain
contract wrappers and access restricting membranes. We evaluate these
designs with respect to usability. We further implement them in the
Firefox SpiderMonkey JavaScript engine and evaluate the impact of
transparent proxies on benchmark performance. 





\paragraph*{Overview and contributions}

Section~\ref{sec:proxies} introduces the JavaScript Proxy API, explains the membrane pattern, and
sketches the implementation of a contract system based on proxies. Section~\ref{sec:usecases} discusses
different use cases of proxies and assesses them with respect to the requirements on proxy transparency.
Section~\ref{sec:break} contains an in-depth discussion of the programmer's expectation from an equality
operation and how it would be affected by the design of a proxy API.  
Section~\ref{sec:alternatives} presents alternative designs to obtain transparency. \textbf{We
  present two novel designs for transparent proxies that do not impede the implementation of advanced
proxy management. }
In Section~\ref{sec:implementation}, we describe how \textbf{we implement these two designs in
the Firefox JavaScript engine (interpreter and baseline JIT compiler)}.
In Section~\ref{sec:evaluation}, \textbf{we demonstrate that our modification to the
JavaScript engine does not affect benchmark performance}. Furthermore, \textbf{we present evidence that
the danger of interference for a contract implementation based on opaque proxies is real}: it
arises in instrumented benchmark programs, not just in artificially constructed examples.
Section~\ref{sec:related_work} discusses related work and
Section~\ref{sec:conclusion} concludes. 

The appendix~\ref{sec:algo-equality}, \ref{sec:algo-identity}, and \ref{sec:algo-equals} contains 
detailed descriptions of the two algorithms used in our implementation.

The implementation of the JavaScript engine with transparent proxies is available in a Github
repository\footnote{\url{https://github.com/sankha93/js-tproxy}}.


\section{Proxies, Membranes, and Contracts}
\label{sec:proxies}

This section introduces the JavaScript Proxy API and presents two
typical applications of proxies:
membranes
that regulate access to an object network and contracts 
that check assertions on the values manipulated by a program.

\subsection{Proxies}

A proxy is an object intended to be used in place of a \emph{target
  object}. As the target object may also be a proxy, 
we call the unique non-proxy object that is transitively reachable through a
chain of proxies the \emph{base target} for each proxy in this chain. The behavior of a
proxy is controlled by a \emph{handler object}, which may modify the original
behavior of the target object in many respects.
A typical use case is to have the handler mediate access to the target
object, but in JavaScript the handler has a full range of intercession
facilities that we only touch on.

The JavaScript Proxy API \cite{CutsemMiller2010} contains a proxy constructor 
that takes the designated target object and a handler object:
\begin{lstlisting}[name=proxy]
var target = { /* some properties */};
var handler = { /* trap functions... */ };
var proxy = new |Proxy| (target, handler);
\end{lstlisting}


The handler object contains optional trap functions that are called when the
corresponding operation is performed on the proxy. Operations like
property read, property assignment, and function 
application are forwarded to the corresponding trap. The trap function may implement
the operation arbitrarily, for example, by forwarding the operation to the
target object. The latter is the default functionality if the trap is
not specified.

Performing an operation like
property get or property set on the proxy object results in a meta-level
call to the corresponding trap on the handler object. For example, the property get operation 
\lstinline{proxy.x} invokes 
\lstinline{handler.get(target,'x',proxy)} and the property set operation
\lstinline{proxy.y=1} invokes the trap
\lstinline{handler.set(target,'y',1,proxy)},
if these traps are present. 
%
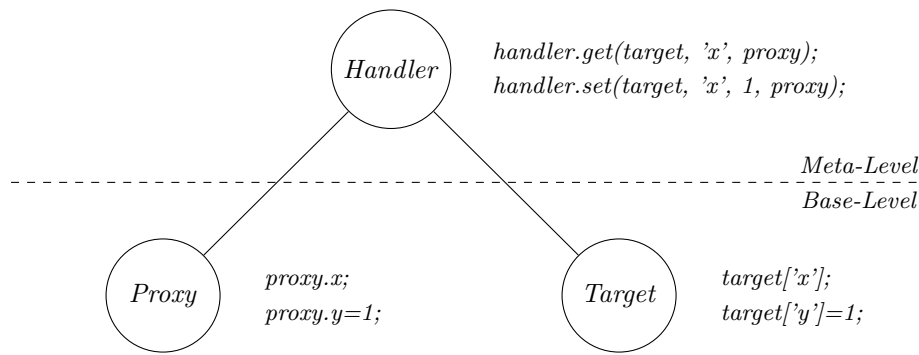
\begin{figure}[t]
  \centering
\tikzstyle{level 0}=[level distance=4cm, sibling distance=2cm]
\tikzstyle{level 1}=[level distance=3cm, sibling distance=6cm]
\tikzstyle{level 2}=[level distance=3cm, sibling distance=4cm]
\tikzstyle{level 3}=[level distance=1cm, sibling distance=2cm]
\tikzstyle{level 4}=[level distance=1cm, sibling distance=2cm]
  \begin{tikzpicture}[-,level/.style={sibling distance=180mm/#1}]
    \node [draw, solid, circle, minimum width=1.5cm] (handler) {\textit{Handler}}
    child {%
      node [draw, solid, circle, minimum width=1.5cm] (proxy) {\textit{Proxy}}
    }
    child {%
      node [draw, solid, circle, minimum width=1.5cm] (target) {\textit{Target}}
    };

    \node[draw=none, anchor=west, right=1cm] at (proxy) {
      \begin{tabular}{l}
        \lstinline!proxy.x;!\\
        \lstinline!proxy.y=1;!
      \end{tabular}
    };

    \node[draw=none, anchor=west, right=1cm] at (handler) {
      \begin{tabular}{l}
        \lstinline!handler.get(target, 'x', proxy);!\\
        \lstinline!handler.set(target, 'x', 1, proxy);!
      \end{tabular}
    };

    \node[draw=none, anchor=west, right=1cm] at (target) {
      \begin{tabular}{l}
        \lstinline!target['x'];!\\
        \lstinline!target['y']=1;!
      \end{tabular}
    };

    \draw [dashed] (-5,-1.5) -- (7, -1.5)
    node[above, pos=0.93]{\small\textit{Meta-Level}}
    node[below, pos=0.93]{\small\textit{Base-Level}}
    ;
  \end{tikzpicture}

  \caption{Example of a proxy operation}
  \label{fig:proxy_pattern}
\end{figure}
Figure~\ref{fig:proxy_pattern} illustrates this situation with a
handler that forwards all operations to the target. 

However, a handler may redefine or extend the semantics of an operation
arbitrarily. For example, a handler may forward a property access to its target
object only if the property is not locally present. The following example
demonstrates a handler that implements a copy-on-write policy for its target by
intercepting all write operations and serving reads on them locally.
Thus, reading 
\lstinline{target.a} at the end may return a different value than \lstinline{42}.
\begin{lstlisting}[name=proxy]
function makeHandler() {
  var local = {};
  return {
    get: function(target, name, receiver) {
      return (name in local) ? local[name] : target[name];
    },
    set: function(target, name, value, receiver) {
      return local[name]=value;
    }
  };
}
var child = new |Proxy| (target, makeHandler());
child.a = 42; // does not change target
\end{lstlisting}

Proxy and handler objects are based on the JavaScript Proxy API
\cite{CutsemMiller2010,CutsemMiller2013}, which is part of the
JavaScript draft standard ES6. This API is implemented in Firefox since version 18.0 and in
Chrome V8 since version 3.5.

JavaScript's proxies are \emph{opaque}: each proxy object has
its own identity different from all other (proxy) objects. The proxy identity is
observable with the JavaScript equality operators \lstinline{==} and
\lstinline{===}: When applied to two objects, both operators compare object
identities.\footnote{If one argument has a primitive type, \lstinline{==}
attempts to convert the other argument to the same primitive type, whereas
\lstinline{===} returns false if the types are different. If both arguments are
objects, then both operators do the same.
}

The following example (which continues the preceding code fragment)
illustrates this behavior. Comparing distinct proxies 
returns false even though the underlying target is the same. 
Similarly, the target object is different from any of its proxies. 
\begin{lstlisting}[name=proxy]
var proxy2 = new |Proxy| (target, handler);
(proxy==proxy2); // false
(proxy==target); // false
\end{lstlisting}

We already mentioned in the introduction that equality cannot be
trapped. While there are good reasons for this design decision
\cite{CutsemMiller2013}, we mention in passing that it would be hard
to design and implement an efficient equality trap because equality is
a binary method. 


\subsection{Membranes}
\label{sec:membrane}

\tikzstyle{level 0}=[level distance=2cm, sibling distance=2cm]
\tikzstyle{level 1}=[level distance=2cm, sibling distance=2cm]
\tikzstyle{level 2}=[level distance=1cm, sibling distance=2cm]
\tikzstyle{level 3}=[level distance=1cm, sibling distance=2cm]
\tikzstyle{level 4}=[level distance=1cm, sibling distance=2cm]
\tikzset{>=latex}
\begin{figure}[t]
  \centering

  \begin{tikzpicture}[%
    edge from parent/.style={draw,-latex},
    level/.style={sibling distance=180mm/#1}]

    \node [left=1.5cm, draw, solid, circle] (root) {\textit{P1}}
    child {%
      node [draw, solid, circle] (rootl) {\textit{P2}}
    }
    child [edge from parent/.style={draw=none}] {%
      node [draw=none, solid, circle] (roodr) {}
      child {%
        node [draw, solid, circle] (rootrr) {\textit{P3}}
      }
    };
    \node [right=1.5cm, draw, draw, circle] (mroot) {\textit{T1}}
    child {%
      node [draw, solid, circle] (mrootl) {\textit{T2}}
    }
    child [edge from parent/.style={draw=none}] {%
      node [draw=none, solid, circle] (mrootr) {}
      child {%
        node [draw, solid, circle] (mrootrr) {\textit{T3}}
      }
    };    
    \path[draw,->] (root) -- (rootrr);
    \path[draw,->] (rootl) -- (rootrr);

    \path[draw,->] (mroot) -- (mrootrr);
    \path[draw,->] (mrootl) -- (mrootrr);

    \path (root) -- (rootl) node[midway, left=1pt, pos=0.3]{\textit{x}};
    \path (root) -- (rootrr) node[midway, right=1pt, pos=0.3]{\textit{y}};
    \path (rootrr) -- (rootl) node[midway, right=7pt, pos=1]{\textit{z}};

    \path (mroot) -- (mrootl) node[midway, left=1pt, pos=0.3]{\textit{x}};
    \path (mroot) -- (mrootrr) node[midway, right=1pt, pos=0.3]{\textit{y}};
    \path (mrootrr) -- (mrootl) node[midway, right=7pt, pos=1]{\textit{z}};

    \node[dashed,rounded corners=7, draw,fit={(mroot) (mrootl) (mrootr) (mrootrr)}] {}; 

    \begin{pgfonlayer}{background}
      \path[draw,loosely dotted] (root) -- (mroot);
      \path[draw,loosely dotted] (rootl) -- (mrootl);
      \path[draw,loosely dotted] (rootrr) -- (mrootrr);
    \end{pgfonlayer}

  \end{tikzpicture}

  \caption{Property access through membrane}
  \label{fig:membrane_pattern}
\end{figure}
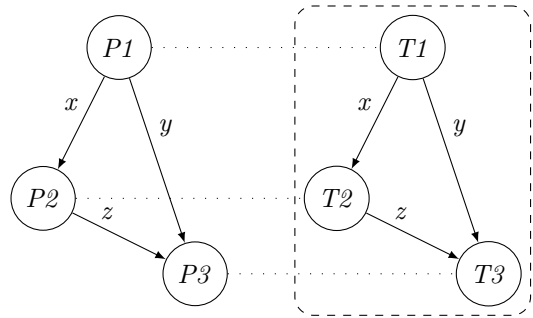

A \emph{membrane} is a regulated communication channel between an
object and the rest of the program. It ensures that all objects
reachable from an object behind the membrane are also behind the
membrane. Figure~\ref{fig:membrane_pattern} shows a
membrane (dashed line) around targets \textit{T1}, \textit{T2}, and
\textit{T3} implemented by the wrapper objects \textit{P1},
\textit{P2}, and \textit{P3}.  
Each property access through the wrapper (e.g., \lstinline{P1.x})
returns a wrapper for \lstinline{T1.x}, which is created on demand. After
installing the membrane, no \emph{new} direct references to target
objects behind the membrane become available. 
%
%
This mechanism may be used to revoke all references to an object
network at once or to
enforce write protection on the objects behind the membrane
\cite{CutsemMiller2010,Miller2006}. 

An \emph{identity preserving membrane} is a membrane that furthermore
guarantees that no target object has more than one proxy. Thus, proxy
identity outside the membrane reflects target object identity
inside.
That is, if \lstinline{T1.x.z===T1.y}, then also
\lstinline{P1.x.z===P1.y}. Figure~\ref{fig:membrane_pattern} depicts
such an identity preserving membrane.

Both kinds of membrane may be implemented with the JavaScript
Proxy API and a weak map that associates target objects with their proxies \cite{CutsemMiller2010}.


%
\subsection{Contracts}
\label{sec:contracts}

Dynamically checked software contracts lie at the core of Meyer's
\emph{Design by Contract}{\texttrademark} methodology \cite{Meyer1988}.
A contract specifies the interface of a software component by stating
obligations and benefits for the component's implementors and users. Contracts
state invariants for objects as well as preconditions and postconditions for
functions and methods.
Contracts are particularly important for dynamically typed languages as these
languages do not provide static guarantees beyond memory safety. For
such languages, contract systems are indispensable tools to create maintainable software.

A contract may specify a property of a value. In many cases, a simple
assertion suffices, but many interesting properties of functions and
objects cannot be checked immediately. For example, the contract
\textit{Num}$\to$\textit{Num} 
expresses that a function maps a number to a number. It can only be
checked when the function is called: the caller must provide a number argument and
then the result must be a number, too. Similarly, a check that a
certain object property must always be assigned a number can only be checked
when actually setting the property. 



We call such contracts on functions and objects \emph{delayed
contracts}, because their assertion never signals a contract
violation immediately. 
The standard implementation of a delayed contract is by wrapping the
function/object in a proxy. The proxy's handler implements traps to 
mediate the use of the function/object and to assert its contract when the
function is called or the object is read or written to. 

The following example sketches the implementation of a contract assertion for
\textit{Num}$\to$\textit{Num}. It installs a proxy where the handler
supplies an \lstinline{apply} trap that gets invoked when the proxy is
called as a function. The arguments to \lstinline{apply} are the
target object, the \lstinline{this} object, and an array containing
the arguments.

\begin{lstlisting}[name=proxy]
var handler = { 
  apply: function(target, thisArg, argsArray) {
    if (typeof argsArray[0] !== 'number') { throw new Exception (); }
    var result = target.apply(thisArg, argsArray);
    if (typeof result !== 'number') { throw new Exception (); }
    return result;
  }
};
var addOne = function (x) { return x+1; };
var addOneNN = new |Proxy| (addOne, handler);
\end{lstlisting}

Thus, the monitored function \lstinline{addOneNN} is implemented by a
proxy with a suitable handler. The contract systems Contract.js
\cite{Disney2013:contracts} and TreatJS \cite{KeilThiemann:TreatJS}
are both implemented in this way.
As JavaScript does not support ``proxification'' (i.e., the
transformation of an arbitrary object into a proxy), it is conceivable
that \lstinline{addOne}, the original object, and \lstinline{addOneNN},
the proxy, are both accessible in the same execution environment.


\section{Opacity vs. Transparency for Proxies}
\label{sec:usecases}

In this section, we question
whether JavaScript proxies need be opaque by considering various use cases for
proxies and evaluating whether they could be served equally well with
transparent proxies, where equality is defined as equality of base
targets.

\subsection{Use Case: Object Extension}
\label{sec:use-case-object-extension}

A common use case of proxies is to extend or redefine the semantics of
particular operations on objects. For example, a handler may throw an
exception instead of returning \lstinline{undefined}, it may redirect different 
operations to different targets (for example to store changes locally or to
implement placeholders), it  may log or trace operations, or it may notify observers.

In this case, using the proxy may lead to a completely different
outcome than using the target object. Thus, proxy and target object should not be
confused.

\subsection{Use Case: Access Control}
\label{sec:use-case-access-control}

Revocable references are the motivating use case for membranes
\cite{CutsemMiller2010,Miller2006}. 
Instead of passing a target object to an untrusted piece of code, 
the idea is to pass its proxy wrapped in a protecting membrane.
Once the host application deems that the untrusted code has finished its
job, it revokes the reference which detaches the proxy from its target.
The membrane extends this detachment recursively
to all objects reachable from the original target.

Opaque proxies are suitable for implementing membranes as well as
their identity preserving variant. However, transparent proxies would
work just as well, because
the host application only sees original objects whereas
the untrusted code only sees proxies. Furthermore, the implementation of revocable
references and membranes ensures that there is at most one proxy for each
original object. If an execution environment is compartmentalized like
this, then each compartment has a consistent view with unique
object (or proxy) references, regardless whether proxies are opaque or
transparent. In fact, with transparent proxies, a membrane is always
identity preserving, the weak map only improves the space efficiency.

\subsection{Use Case: Contracts}
\label{sec:use-case-contracts}

Proxies implement contracts in Racket
\cite{StricklandTobinHochstadtFindlerFlatt2012} and in JavaScript
\cite{Disney2013:contracts,KeilThiemann2013-Proxy,KeilThiemann:TreatJS}.
During maintenance, a programmer may add contracts to a program as
understanding improves. To systematically investigate a program in this
way, the addition of a new contract must not change
a program execution that respects the contract already. In this scenario, the
program executes in a mix of original objects and proxy objects. Furthermore,
there may be more than one proxy (implementing different contracts) for the same
target. If introducing proxies affected the object identity, then some equality
comparisons on objects (\texttt{eq?} in Racket and \lstinline{==}, \lstinline{!=},
\lstinline{===}, or \lstinline{!==} in JavaScript) would flip their outcome,
thus changing the semantics.

Our experimental evaluation (Section~\ref{sec:object_comparisons}) considers a typical
program understanding and maintenance scenario where a programmer inserts
assertions/contracts to document and validate his/her understanding of the
program. We find that mixed (proxy vs.\ non-proxy) object comparisons occur in
realistic programs. 

Similar incidents where observed when maintaining Racket's preferences
framework. Registering a callback with the framework wrapped the
callback in a contract before storing it in an \texttt{eq?}-based weak
hash table. Because there were no further references to the wrapper,
the weak table released it on the next garbage collection. Thus, the
callback disappeared mysteriously, leading to unwanted
behavior\footnote{Personal communication with Robby Findler, February
  2015.}. This problem is evidence that such mixes occur in real
situations and also in a system where contracts are aligned with
module boundaries.

Thus, we see strong evidence that unintended mixing is hard to avoid even in a
well-designed system. If a module has a higher-order interface, then a function
passed as a parameter may capture an un-proxied version of an object that is
also passed as a regular parameter. For example, let \lstinline{T} be
some delayed contract and consider the module interface
\begin{lstlisting}[name=proxy]
// foo : (T $\rightarrow$ boolean, T) $\rightarrow$ boolean*'\label{line:contract-on-foo}'*
\end{lstlisting}
so that \lstinline{foo} carries a wrapper that asserts this contract. The function accepts two
parameters, the first of which is a function. An external caller may use \lstinline{foo} as follows:
\begin{lstlisting}[name=proxy]
var x = { /* some object */ };
var f = function (y) { return x===y; };*'\label{line:body-of-f}'*
foo(f, x);
\end{lstlisting}
The call to \lstinline{foo} wraps \lstinline{f} and \lstinline{x} in
the respective contract wrappers for \lstinline{T $\to$ boolean} and
\lstinline{T}. Unfortunately, wrapping the function \lstinline{f} does
not affect the free variable \lstinline{x} in its environment. 


Now consider the following contract-abiding implementation of \lstinline{foo}:
\begin{lstlisting}[name=proxy]
function foo (f, y) { return f (y); }
\end{lstlisting}
Inside of \lstinline{foo}, \lstinline{y} is wrapped in the
\lstinline{T} contract and applying \lstinline{f} (wrapped in
\lstinline{T $\to$ boolean}) may wrap it one more time in a
\lstinline{T}. Thus, in the body of \lstinline{f}
(line~\ref{line:body-of-f}), \lstinline{x} is unwrapped and
\lstinline{y} is the same object wrapped at least once in \lstinline{T}. Thus, assuming opaque
proxies, \lstinline{x===y} yields \lstinline{false}, which is different from the result before
installing the contract on \lstinline{foo} (line~\ref{line:contract-on-foo}). 

Thus, the existing implementations of higher-order contracts for JavaScript are
prone to interfere with the semantics.  The situation is similar for Racket.
Racket's chaperones and impersonators are opaque
because they may be distinguished from their target and from one another using
\texttt{eq?}.  This choice is legitimized by pointing out that 
the preferred equality test in Racket is the \texttt{equal?} function that
compares two values for \emph{structural equality}, a non-trivial
functionality that is not available out-of-the-box in JavaScript (cf.\
\cite{AdamsDybvig2008}). Clearly, chaperones and impersonators are
transparent with respect to \texttt{equal?}.
The paper on chaperones and impersonators
\cite{StricklandTobinHochstadtFindlerFlatt2012} further acknowledges that
``wrappings do affect the identity of objects, as compared with
JavaScript's \lstinline{===} or Racket's \texttt{eq?} comparisons'',
but remarks that Racket programmers rely much less on object
comparison than JavaScript programmers. Indeed, the paper's formalization
includes \texttt{equal?}, but not \texttt{eq?}.

\subsection{Assessment}
\label{sec:assessment}

Neither the opaque nor a transparent proxy implementation can be labeled as
right or wrong without further qualification. Each is appropriate for 
particular applications and may lead to undesirable behavior in other applications.

Opaqueness is required for a proxy that changes the behavior of its target significantly. This case
corresponds to the use case  for impersonators~\cite{StricklandTobinHochstadtFindlerFlatt2012}.

Transparent proxies can safely be used to implement revocable
membranes as well as for other applications that guarantee compartmentalized
execution (where proxies and targets never meet). Their use simplifies the implementation of the
identity preserving membrane because the weak map from targets to their proxies may be
elided. However, the price for this elision is increased space usage for multiple wrappers for the same
target.  It must be weighed against the time taken to administer
a weak target-to-proxy mapping.

Transparency is required for proxies that implement contract wrappers to avoid
the interference with the normal program
execution pointed out in the introduction and in 
Section~\ref{sec:use-case-contracts}. To avoid such interference with
opaque proxies, a programmer would have to guarantee that contracts
are only ever applied to unique object references or that references to
the target object do not escape. Such a
guarantee may be established by only applying contracts to objects as
they are created or by static analysis. However, the
former is an unrealistic assumption and the latter severely limits the
freedom that developers want to obtain by using contracts in a dynamically typed language: rather than
submitting to a type system, they must submit to a uniqueness
or an escape analysis. 

A similar case can be made for further wrapper-based programming
patterns. For example, a wrapper that records all operations
performed on an object reference can be very helpful while
debugging. Clearly, such a wrapper must be indistinguishable from its
target object. 

It is also clear that the behavior of equality is not something that should be
left to the whim of the programmer. For example, equality on objects should be
an equivalence relation, which means that the equality operations \lstinline{==}
and \lstinline{===} must not be trapped \cite{CutsemMiller2013}.

Thus, the current state of affairs in JavaScript is fully justified, but it is
not well suited to implement contract systems. To obtain some insight
into a proxy design suitable for implementing contracts, we first
explore the important issue of equality and then consider some designs
and assess the suitability with respect to the use cases outlined in
this section.


\section{Invariants for Equality}
\label{sec:break}

What does a programmer expect from an equality function on objects?
Let's consult the language references for Racket, Java, and JavaScript for cues.

Racket inherits its hierarchy of equality operators from Scheme:
\texttt{equal?}, \texttt{eqv?}, and \texttt{eq?}. The Scheme report \cite{R6RS-cup}
specifies them as follows: ``An equivalence predicate is the computational
analogue of a mathematical equivalence relation;
it is symmetric, reflexive, and transitive. Of the equivalence
predicates described in this section, \texttt{eq?} is the finest
or most discriminating, \texttt{equal?} is the coarsest, and \texttt{eqv?}
is slightly less discriminating than \texttt{eq?}.'' The intuition is
that \texttt{equal?} implements structural equality, \texttt{eq?}
implements pointer equality, and \texttt{eqv?} is a compromise between
the expensive structural equality\footnote{It is supposed to work on
  cyclic structures, too. \cite{AdamsDybvig2008}} and 
pointer equality, which may distinguish different boxings of the same number. 

There are further differences between \texttt{eq?} and
\texttt{equal?}. The function \texttt{eq?} is guaranteed to run in
$O(1)$, whereas there is no known implementation of \texttt{equal?} that runs in less than linear
time. Furthermore, \texttt{equal?} is
not stable in the sense that \texttt{(equal?\ x y)} may hold at some point during execution, but this
equality may be destroyed by subsequent assignments in the program. In contracts,
\texttt{(eq?\ x y)} does not change as the program proceeds.


Java has an \texttt{equals} method in \texttt{java.lang.Object}. The
documentation of this class
reads:\footnote{\url{http://docs.oracle.com/javase/7/docs/api/java/lang/Object.html}}
``The equals method implements an equivalence relation on non-null
object references: \dots'' It also asks that repeated invocations with the same arguments behave
consistently in that they always return the same answer. And finally:
``The equals method for class Object implements the most
discriminating possible equivalence relation on objects; that is, for
any non-null reference values \texttt{x} and \texttt{y}, this method returns \texttt{true} if and
only if \texttt{x} and \texttt{y} refer to the same object (\texttt{x == y} has the value
\texttt{true}).'' About the \texttt{==} operator, the JLS
says:\footnote{%
  \url{http://docs.oracle.com/javase/specs/jls/se7/html/jls-15.html} in 15.21
} ``The equality operators are commutative \dots'' and then ``the result
of \texttt{==} is \texttt{true} if the operand values are both
\texttt{null} or both refer to the same object or array; otherwise,
the result is \texttt{false}.'' 
Regarding stability and execution time, the \texttt{==} operator is stable and runs in $O(1)$. The
\texttt{equals} methods is recommended 
to be implemented in a stable way, but this restriction is not enforced. No bounds on the execution
time are prescribed and none can be given.

The ECMAScript specification \cite[Section 11.9]{ecmascript2009} does not mention
algebraic properties of the equality operation, but rather contains pseudocode for the abstract equality comparison
algorithm, which underlies the \lstinline{==} operator. This algorithm implements an operation, which
is symmetric, but neither reflexive nor transitive: It is not reflexive because \lstinline{NaN==NaN}
is \lstinline{false} and it is not transitive because \lstinline{new String("foo")=="foo"} and
\lstinline{"foo"==new String("foo")}, but \lstinline{new String("foo")==new String("foo")} is
\lstinline{false} due to unfortunate interaction with type conversions. Restricted 
to object arguments the algorithm says: ``Return true if x and y refer to the same
object. Otherwise, return false.'' The strict equality comparison algorithm (in Section 11.9.6),
which specifies the \lstinline{===} operator, is not reflexive because of \lstinline{NaN}, but appears to
be symmetric and transitive. Thus, it is at least a partial equivalence relation.
Restricted to objects, both equalities are equivalences if that is implied by sameness.

JavaScript's \lstinline{==} operator is not stable: Consider a comparison between an object and a
string and then change the \lstinline{toString} method of the object. However, \lstinline{===} is
stable because it does not involve type conversion. Restricted to object arguments, both
equalities are stable.


The least common denominator for an object equality appears to be that \textbf{equality must be a
  stable equivalence relation}. Often, object equality is explained by alluding to the ``sameness''
of objects, which is left to further interpretation. In particular, the JLS explicitly mentions that
\texttt{==} is overly discerning for \texttt{String} 
objects and none of the language specifications addresses proxy objects. 

Some proponents of opaque proxies ask that equality should be an observational equivalence. That is,
for any conditional of the form
\begin{lstlisting}[name=proxy]
if (a === b) Statement
\end{lstlisting}
(where \lstinline{a} and \lstinline{b} are variables) it should be possible to substitute
\lstinline{b} for \lstinline{a} (but respecting lexical binding rules) in \lstinline{Statement}
without changing the semantics. 
However, in the presence of JavaScript's \lstinline!with (head) { ... body ... }! statement inside
\lstinline{Statement} further qualifications are needed. As \lstinline{with} extends the lexical
environment by placing \lstinline{head} on top of the scope chain while executing \lstinline{body},
a property of \lstinline{head} may shadow any variables in scope including \lstinline{a}.
Thus, the substitution must not extend to the body of a \lstinline{with} statement.

If the above conditional appears in the \lstinline{body} of a
\lstinline!with (head)  { ... body ... }! statement, then essentially all bets are off.  The object
\lstinline{head} may define \lstinline{a} or \lstinline{b} via a getter function or via a proxy handler, which may be
impure and return different results on each access. Alternatively, \lstinline{a} and
\lstinline{b} may be changed by simple assignment to \lstinline{head}.

Furthermore, the \lstinline{Statement} may contain a call to \lstinline{eval}. As this call is
executed in the lexical environment of its call site, its execution may perform an assignment to
\lstinline{a} or \lstinline{b}. As the call to \lstinline{eval} may be performed indirectly by
storing \lstinline{eval} in a different variable or object property, essentially every function or
method call may assign to \lstinline{a} or \lstinline{b}. 

While ES5 strict mode abolishes the \lstinline{with} statement and severely restricts the use of
\lstinline{eval}, a call to \lstinline{eval} may still assign to variables in scope at the
point of the call. For example, the following function \lstinline{f42} always returns \lstinline{42}
because the \lstinline{eval} assigns to an existing local variable.
\begin{lstlisting}
  function f42(x) {
    "use strict";
    eval("x = 42'');
    return x;
}
\end{lstlisting}

Thus, arriving at a practically useful definition of observational equivalence in
JavaScript is quite intricate; perhaps too intricate to be a useful reasoning tool for programmers.
Moreover, we are not aware of a language definition that requires its equality operator to be an
observational equivalence.

What is the take-home message from this discussion with respect to the question whether a proxy
implementation should be opaque or transparent with respect to equality? We believe that most
programming patterns using equality only expect a stable equivalence relation, which is easy to
implement for transparent proxies, as we show in Section~\ref{sec:transparent-proxies}. However,
disregarding the problems with the \lstinline{with} statement and \lstinline{eval}, a
weak version of observational equivalence is certainly desirable: For any conditional of the form
\begin{lstlisting}[name=proxy]
if (a === b) Statement
\end{lstlisting}
if we substitute \lstinline{b} for \lstinline{a} in \lstinline{Statement}, then either the semantics
of \lstinline{Statement} does not change or \lstinline{Statement} raises an exception. 

However, a transparent proxy without further restrictions would not even fulfill weak observational
equivalence. As a compromise, the behavioral change of a transparent proxy could be
restricted to implement a \emph{projection}, a point that we come back to in
Section~\ref{sec:observer-proxies}.

\section{Design Alternatives for Proxy Equality}
\label{sec:alternatives}

In this section, we explore some alternatives for designing proxies and equalities and discuss their
suitability for the use cases outlined in Section~\ref{sec:usecases}.


\subsection{Program Rewriting}
\label{sec:program-rewriting}

One way to obtain transparent proxies with an implementation of opaque proxies
is to replace all occurrences of \lstinline{==}, \lstinline{!=}, \lstinline{===}, and
\lstinline{!==} with proxy-aware equality functions.
These functions can be implemented in JavaScript using a weak map from proxies to their target.
The proxy constructor would be extended to maintain this map. It would be possible to treat
some proxies as transparent and the rest as opaque.



This approach preserves the existing behavior and retains the possibility to
distinguish proxies from target objects in library code implementing proxy
abstractions. 
A macro system like SweetJS \cite{DisneyFaubionHermanFlanagan2014} may be used to implement such a
transformation elegantly, alas, SweetJS is currently implemented as an offline transformation and
would need to be extended to deal with \texttt{eval} and dynamic loading.

\subsection{Additional Equality Operators}

Another approach would be to reinterpret the JavaScript equality
operators \lstinline{==} and \lstinline{===} as proxy-transparent and introduce new variants, 
\lstinline{:==:} and \mbox{\lstinline!:===:!,} say, for their opaque cousins (i.e., the current
implementations of  \lstinline{==} and \lstinline{===}). The former operators are used in application code 
whereas the implementation of proxy abstractions would make use of the opaque
operators where needed.



No code transformation is required with this approach. However, it is not clear
how to ensure that application code does not use the opaque operators. It is not
even clear if it \emph{should not} use them. While proxy abstractions can be
implemented, the distinction between application and library seems too brittle.

Given both operations, application code can test if two objects are in a proxy relation with the
same target:

\begin{lstlisting}[name=proxy]
((x === y) != (x :===: y)); // true, iff x is in a proxy relation to y
\end{lstlisting}

Furthermore, the implementation would either have to use macros, which gives rise to the problems
discussed in Section~\ref{sec:program-rewriting}, or it would have to be implemented in the
JavaScript engine, where it requires changes starting in the parser. This point makes it unlikely
that such a proposal would be adopted by the 
community. Moreover, a proliferation of equality operations is confusing for developers as there are
already three different kinds of equality in JavaScript: besides equality and strict equality, there is a third equality
that fixes reflexivity when comparing \lstinline{NaN} values.





\begin{lstlisting}[float,label={lst:getidentityobject},caption={Pseudo code for GetIdentityObject}]
function GetIdentityObject(obj) {
  while(isProxy(obj) && isTransparentProxy(obj)) {
    obj = getProxyTargetObject(obj);
  }
  return obj;
}
\end{lstlisting}

\subsection{Transparent Proxies in the VM}
\label{sec:transparent-proxies}

We already discussed that trapping the equality operation is not appropriate.
As an alternative, we implement transparent proxies as an extension of a JavaScript VM
(cf. Section~\ref{sec:implementation}) and provide different constructors for transparent and for
opaque proxies.   
This extension provides a different kind of proxy object
on which equality comparison behaves differently. Before testing reference identity as the
last step in a comparison of two objects, the
equality comparison calls a new internal function \lstinline{GetIdentityObject} (see
Listing~\ref{lst:getidentityobject}) that computes the base target of an object.
For a non-proxy object, the function returns its argument. For a proxy object,
\lstinline{GetIdentityObject} checks whether the proxy is transparent. If that check fails, then
\lstinline{GetIdentityObject} returns the reference to the current proxy object.
Otherwise, it iteratively performs the same checks on the proxy's
target. For consistency, the \lstinline{GetIdentityObject} method also needs to
be called in other computations that depend on object identity. One example is the
WeakMap abstraction of the ES6 draft standard.

This design enables both scenarios described in
Sections~\ref{sec:use-case-access-control} and~\ref{sec:use-case-contracts}.
It also guarantees that equality (on objects) is an
equivalence relation.


Transparent proxies need special attention because there are abstractions that
require to test whether a reference is a (transparent) proxy. For example, the implementation of access permissions
contracts \cite{KeilThiemann2013-Proxy} extracts the current permission from a
proxy to construct a new proxy with an updated permission. This introspection improves the
efficiency of the implementation; its absence would lead to long, inefficient chains of
proxy objects.

Thus, for implementing proxy abstractions it is useful to be able to break the
transparency. We propose to use secret tokens for this purpose. A token
(just an object in JavaScript) stands for a transferable right to perform a
particular operation. We attach the token object to a proxy by making it an extra argument to
the constructor of transparent proxies, \lstinline{TransparentProxy}, say. Being a standard object,
the token can be hidden in the scope of the function that wraps objects.

\begin{lstlisting}[name=proxy]
var wrap = (function() {
  var token = {};
  return function(target, handler) {
    return new TransparentProxy(target, handler, token);
  };
  // further operations on wrappers
})();
\end{lstlisting}

Later on, the token can be used to make the transparent proxy visible for
equality operations. To this
end, we need an equality operation \lstinline{Object.equals} that takes the
token as a third (optional) parameter.  
The following snippet demonstrates this operation in action.

\begin{lstlisting}[name=proxy]
var token = {};
var target = { ... };
var proxyA = new TransparentProxy (target, handlerA, token);
var proxyB = new TransparentProxy (target, handlerB, token);
target == proxyA; // true
proxyA == proxyB; // true
Object.equals(target, proxyA, token); // false: token reveals proxy identity
Object.equals(proxyA, proxyB, token); // false
Object.equals(target, proxyA, { /* some other object */ }); // true
Object.equals(proxyA, proxyB, { /* some other object */ }); // true
\end{lstlisting}

Weak maps and other internal data structures that depend on object equality may
be extended with tokens in the same way.

From different point of view, the tokens assign transparent proxies to distinct realms.
Thus, instead of passing tokens one could use object
capabilities to create proxies in a particular realm and to create an equality
function that only reveals proxies for that realms\footnote{%
We would like to thank the anonymous ECOOP\,2015 reviewer who suggested this
elegant solution.}.
A realm constructor may be implemented in JavaScript on top of our token-based implementation.
\begin{lstlisting}[name=proxy]
TransparentProxy.createProxyConstructor = function() {
  var token = {};
  var equals = function (x, y) {
    return Object.equals(x, y, token);
  }
  var Constructor = function(target, handler) {
    return new TransparentProxy(target, handler, token);
  }
  return {Constructor:Constructor, equals:equals};
}
\end{lstlisting}
The realm constructor returns a new transparency realm represented by an object that consists of a
fresh constructor for transparent proxies (named 
\lstinline{Constructor}) and an \lstinline{equals} function revealing proxies of that realm.
\begin{lstlisting}[name=proxy]
var realm = TransparentProxy.createProxyConstructor();
var proxy1 == realm.Constructor(target, handler);
var proxy2 == realm.Constructor(target, handler);
\end{lstlisting}
The proxies \lstinline{proxy1} and \lstinline{proxy2} are transparent with respect to equality unless someone uses the
\lstinline{realm.equals} method.
\begin{lstlisting}[name=proxy]
proxy1 === proxy2; // true
Object.equals(proxy1, proxy2); // true
realm.equals(proxy1, proxy2); // false
\end{lstlisting}
Here, \lstinline{===} and \lstinline{Object.equals} return \lstinline{true} and
do not reveal proxies. However, the \lstinline{realm.equals} function is a
capability that represents the right to reveal proxies of that realm.
Realm-aware weak maps and other internal data structures can be created in same
way.

\subsection{Observer Proxies}
\label{sec:observer-proxies}

To implement contracts, transparent proxies need not be fully general. It would be sufficient if
transparent proxies would be restricted to \lstinline{Observer}s that implement a projection: they would either
return a value identical to the value that would be returned from the target
object (this should also include the same side effects) or that restricts the
behavior of the target object. An \lstinline{Observer} can cause a program
to fail more often, but otherwise it would behave in the same way as if no
observer were present.

A similar feature is provided by Racket's
chaperones~\cite{StricklandTobinHochstadtFindlerFlatt2012}. A chaperone is a proxy that either
returns an identical value, returns a chaperone of this value, or throws an exception. This
restricted kind of proxy is shown to be sufficient to implement a contract system. (Recall that
chaperones are not transparent.)

The following code snippet sketches the implementation of an
\lstinline{Observer} proxy in JavaScript that mimics Racket's chaperone proxy.
The example considers the \lstinline{get} trap only, but other traps can be
implemented in the same way.
\begin{lstlisting}[name=proxy]
function Observer(target, handler) {
  var sbx = new Sandbox(/* some parameters */); *'\label{lst:sbx}'*
  var controller = { 
    /* further traps omitted */
    get: function(target, name, receiver) {
      var result = (trap=handler['get']) ? sbx.call(trap, target, name, receiver) : undefined;
      var raw = target[name];*'\label{lst:raw}'*
      return (result===raw) ? result : raw;*'\label{lst:return}'*
    }
  };
  return new TransparentProxy(target, controller);
}
var target = { /* some object */ };
var handler = {
  get:function(target, name, receiver) {
    return target[name];
  }
};
var observed = Observer(target, handler);
\end{lstlisting}

The constructor starts with instantiating a sandbox (line~\ref{lst:sbx}). The
sandbox is drawn from another contract system for JavaScript,
TreatJS~\cite{KeilThiemann:TreatJS}, that uses membranes and decompilation to implement access
restrictions. Functions execute inside the sandbox without interfering with the normal execution.

The implementation distinguishes between a user specified handler
(\lstinline{handler}) whose traps are evaluated
in the sandbox to guarantee noninterference and the proxy handler
(\lstinline{controller}) which is used to implement the behavior of the observer.
The controller's \lstinline{get} trap first checks the presence of a
\lstinline{get} trap in the user handler, before it evaluates this trap in the
sandbox. Next, it performs a normal property access on the target value. This
step is required to produce the same side effects and to get a reference value
to compare the results. Finally, the reference value is compared with the
result from calling the trap. Line~\ref{lst:return} makes only sense when using
transparent proxies. The user specific trap can return an observer of the
reference value. 

Indeed, the implementation is only correct if one can ensure that
\lstinline{result} is either the \lstinline{raw} value or a transparent proxy
generated by an \lstinline{Observer}. A user specific handler can simply elude
the observers behavior by returning a transparent proxy of the same target but
with a different handler object.
Correctness can be guaranteed by either hiding transparent proxies from the user level
or by using the sandbox to restrict resources access be the handler's trap. 


\subsection{Recommendation}
\label{sec:recommendation}

There is likely no single semantics for object identity that fits the programmers
expectation in all possible contexts. 
A proxy that changes the behavior of its target object significantly needs its own identity and thus
needs to be implemented opaquely.

A contract proxy that only restricts the behavior of the original object, we propose to use a
transparent observer proxy with the design explained in
Section~\ref{sec:transparent-proxies}. For these proxies, \lstinline{===} (and
friends) will be forwarded to the target objects, recursively. An observer proxy
(see Section~\ref{sec:observer-proxies}) limits the possible change of behavior
analogously to chaperones.
Technically, observer proxy weakly simulate the original
objects.


\section{Implementation}
\label{sec:implementation}

We implemented two prototype extensions of the SpiderMonkey JavaScript
engine, one according to the design of Section~\ref{sec:transparent-proxies} and another which
extends the proxy handler by an \lstinline{isTransparent} 
trap that regulates the proxy's transparency. The first prototype implements a new global object
\lstinline{TransparentProxy} that implements the constructor for transparent proxy objects.

Proxies created with \lstinline{new Proxy} in the second prototype are generally opaque, unless they implement an
\lstinline{isTransparent} trap and this trap returns false. Proxies created with
\lstinline{new TransparentProxy} are generally transparent unless they are compared with \lstinline{Object.equal}
using the correct token. These choices guarantee full backwards compatibility.

The implementation is rather tedious because many things have to be implemented three times as
SpiderMonkey consists of an interpreter, the baseline JIT compiler (JaegerMonkey), and the IonMonkey
compiler. Support for transparent proxies has to be added at each level because SpiderMonkey
switches at run time from interpreter to baseline compiler and then to IonMonkey after a sufficient
number of loop iterations. Specifically, the documentation says: ``All JavaScript functions start
out executing in the interpreter [\dots\ which] collects type information for use by the JITs. When
a function gets somewhat hot, it gets compiled with JaegerMonkey. [\dots] When it gets really hot,
it is recompiled with IonMonkey.'' If type information changes, execution falls back all the way to
the interpreter. 

\subsection{JavaScript Interpreter}

To cater for transparent proxies, the interpreter had to be changed in a few places.
\begin{inparaenum}
\item The internal classes \texttt{Proxy} and \texttt{BaseProxyHandler} had to be extended to
  support the new  \lstinline{isTransparent} trap.
\item The comparison operators had to be modified to recognize transparent
  proxies and obtain their base target for comparison,
\item All internal data structures which are connected to the
  identity of objects (in particular, the map, weak map, and set abstractions of the upcoming
  JavaScript standard) had to be modified.
\end{inparaenum}







\subsubsection{JavaScript's Equality Comparison}

JavaScript provides two types of comparison operators. The \emph{strict equality comparison}
(e.g. \lstinline{===}) returns \lstinline{false} if the operands have different types. The
\emph{equality comparison} (e.g. \lstinline{==})
applies type conversion if the operands have different types; then it essentially performs a strict
comparison on the converted values. When comparing two objects \lstinline{x} and \lstinline{y}, both
comparisons behave identically~\cite[Section 11.9.3]{ecmascript2009}:
\begin{quotation}
  1.f. ``Return \lstinline{true} if \lstinline{x} and \lstinline{y} refer to the same object.'' 
\end{quotation}

Thus, this test for sameness of two objects is only one place where the algorithms for equality
comparison and strict equality comparison have to be changed. Our implementation replaces the test
(case 1.f. in equality comparison [11.9.3], case 7. in strict equality comparison [11.9.6]) as follows.
\begin{enumerate}
  \item Let \lstinline{lhs} be the result of calling \lstinline{GetIdentityObject} on \lstinline{x}.
  \item Let \lstinline{rhs} be the result of calling \lstinline{GetIdentityObject} on \lstinline{y}.
  \item Return \lstinline{true} if \lstinline{lhs} and \lstinline{rhs} refer to the same object.
    Otherwise, return \lstinline{false}.
\end{enumerate}


\subsubsection{Getting the Identity Object}

When comparing two objects or when adding an object to a map, transparent
proxies do not use their own identity. To get the right identity for the object
the operation first checks the transparency of the proxy object and transitively obtains its target
object until either an opaque proxy or a native object is reached (cf.\
Listing~\ref{lst:getidentityobject} for the pseudocode). All object comparisons refer to this internal method. 

The actual implementation is slightly more involved, in particular the implementation that supports
the \lstinline{isTransparent} trap (its existence needs to be checked, it needs to be called, and its
results needs to be interpreted). Technical details may be checked in the source
code which is available in a github repository.

\subsubsection{Maps, Sets, and other Data Structures}

After modifying the comparison operators the internal data structures
\lstinline{Map}, \lstinline{Set}, and \lstinline{WeakMap}, which depend on object equality, have
to be adjusted to handle transparent proxies. If \lstinline{target == proxy}
evaluates to \lstinline{true} then
\mbox{\lstinline{map.has(target)==map.has(proxy)}} should also evaluate to
\lstinline{true}.

When adding a new key-value pair to any Map, WeakMap, or Set, the operation
first determines if the key is an object of type \lstinline{Proxy}. If it is a
\lstinline{Proxy}, then the \lstinline{GetIdentityObject} internal method is used to
determine the identity object; hashing takes place with respect to this
identity object, but the original key is stored in the collection along with its
value. A subsequent lookup of the identity object or any proxy with the same identity object returns
the same stored value.
The implementation of the \lstinline{for...in} loop or calling \lstinline{.keys()}
or \lstinline{.entries()} on a map returns the originally added object as key value. 





The example below demonstrates the behavior just described on an empty map object
\lstinline{map}. We first create one transparent and one 
opaque proxy for the same target. The
operation \lstinline{map.set(target, A)} creates a new map entry, whereas the
second one \lstinline{map.set(proxy1, B);} updates this entry. The
third operation \lstinline{map.set(proxy2, C);} creates a new entry, again.
\begin{lstlisting}[name=proxy]
var target = {};
var proxy1 = new TransparentProxy(target, {});
var proxy2 = new Proxy(target, {});
map.set(target, A); // map = [target *'$\color{gray}{\mapsto}$'* A]
map.set(proxy1, B); // map = [target *'$\color{gray}{\mapsto}$'* B]
map.set(proxy2, C); // map = [target *'$\color{gray}{\mapsto}$'* B, proxy2 *'$\color{gray}{\mapsto}$'* C]
\end{lstlisting}

\subsection{Object.equals}

Transparent proxies created with \lstinline{TransparentProxy} are generally
indistinguishable from their base target object and from another transparent
proxy object of the same target.
However,  \lstinline{Object.equals} can be employed to make them distinguishable for
algorithms that implement advanced proxy manipulation. To this end, the constructor stores its
token argument in a slot of each proxy. 

When \lstinline{Object.equals} is called with arguments \lstinline{obj1},
\lstinline{obj2} and an optional argument \lstinline{secret} the following steps
are taken:
\begin{itemize}
  \item If \lstinline{secret} is not present, then return the value of \lstinline{obj1 === obj2}.
  \item If one of \lstinline{obj1} or \lstinline{obj2} is a transparent proxy where the token
    slot matches the \lstinline{secret}, then return \lstinline{true} if \lstinline{obj1} is the same
    object as \lstinline{obj2} (and \lstinline{false}, otherwise).
  \item Otherwise return the value of the transparent comparison \lstinline{obj1 === obj2}.
\end{itemize}

\subsection{JavaScript Baseline Compiler}

The SpiderMonkey Baseline Compiler is the first tier of the JIT compiler. It produces native code for 
JavaScript through stub method calls and optimized inline caches (ICs) for some operations. To adapt 
for changes to the equality (both strict and non-strict) comparison operation, the fallback stub code 
was modified to do a call to the VM and test for equality between the two objects in exactly the same 
manner as in the interpreter.

For the \lstinline{isTransparent} trap implementation
we stop emitting the optimized ICs for object-object comparison, and instead use the 
fallback IC to do a VM call. This will invoke the \lstinline{isTransparent} trap for the proxy and then 
compare with the identity object if it evaluates to \lstinline{true} or it performs the standard 
object-object comparison.

For the \lstinline{TransparentProxy} implementation, we stop emitting the optimized IC for any object-object 
comparison that involves comparison of \lstinline{TransparentProxy} object. We use the fallback stub to do 
a VM call and do a comparison with the identity object, if it involves a \lstinline{TransparentProxy}. Any 
other kind of comparison operations are left unaffected and still take place through the optimized stubs.

\subsection{IonMonkey Compiler}

The IonMonkey optimizing compiler is the final tier of the SpiderMonkey JIT compiler. This has been left 
unmodified, but we give an outline for a future implementation so that generated native code by
IonMonkey can support transparent proxy comparisons. 

For the \lstinline{isTransparent} trap implementation, it will need to load the object into the register 
and test if the object's class is a \lstinline{Proxy} or not. If it is not a proxy, then 
normal fast path for object-object comparison code can be generated. Otherwise execution should stop, do 
a VM call to the \lstinline{isTransparent} trap in the handler object of the proxy, and store the return 
value in a register. If the value is \lstinline{false} then proceed with emitting the usual object comparison 
operation code, otherwise load the identity object of the proxy (i.e., the result of calling 
\lstinline{GetIdentityObject} on the proxy) and replace that value in the equality operation's operand 
register. Then we can proceed with emitting of code for a standard object comparison.

For the \lstinline{TransparentProxy} implementation, the code generation for the equality operation would 
be simpler, as there is no handler trap to be called. We'd have to load the object into the register and 
test if the object's class is a \lstinline{TransparentProxy} or not. If it is not a \lstinline{TransparentProxy}, 
then normal fast path for object-object comparison code can be generated. Otherwise load the result
of calling \lstinline{GetIdentityObject} on the proxy and replace that value in  
the equality operation's operand register. Then we can proceed with emitting of code for a standard object 
comparison.

\subsection{Getting the Source Code}
\label{sec:download}

The implementation of the modified engine is available on the Web\footnote{%
  \url{https://github.com/sankha93/js-tproxy/}}.
The branch \emph{isTransparent-trap}\footnote{%
\url{https://github.com/sankha93/js-tproxy/tree/isTransparent-trap}
} contains the isTransparent handler trap version and branch 
\emph{global-tproxy-object}\footnote{%
  \url{https://github.com/sankha93/js-tproxy/tree/global-tproxy-object}
} contains the implementation of the TransparentProxy object.
A \emph{README} file in each of the respective branches contains the build instructions.

\section{Evaluation}
\label{sec:evaluation}


This section reports our experiences with applying our modified engines to
JavaScript benchmark programs to answer the following research
questions:
\begin{description}
  \item[RQ1] Does the introduction of transparent proxies affect 
    the performance of non-proxy code?
  \item[RQ2] Does a contract implementation based on opaque proxies
    affect the meaning of realistic programs? 
\end{description}

\subsection{Performance Test}
\label{sec:performance}

To answer \textbf{RQ1}, we evaluate the impact of our implementations of transparent proxies on
programs that do not make use of proxies at all. These programs may be affected by our modifications
to the equality comparison algorithms and to the set and map abstraction.

To this end we used benchmark
programs from the Google Octane 2.0 Benchmark
Suite\footnote{\url{https://developers.google.com/octane/}}.
This suite comprises 17 programs that range from
performance tests to real-world web applications (Figure~\ref{fig:performance}),
that is, from an OS kernel simulation to a portable PDF viewer. Each program
focuses on a special purpose, for example, function and method calls, arithmetic
and bit operations, array manipulation, JavaScript parsing and compilation.

Octane reports its results in terms of a score. The Octane
FAQ\footnote{\url{https://developers.google.com/octane/faq}} explains the score
as follows: ``\emph{In a nutshell: bigger is better. Octane measures the time a
  test takes to complete and then assigns a score that is inversely proportional
to the run time.}'' The constants in this computation are chosen so that the
current overall score (i.e., the geometric mean of the individual scores)
matches the overall score from earlier releases of Octane and new benchmarks are
integrated by choosing the constants so that the geometric mean remains the
same. The rationale is to maintain comparability.

\subsubsection{The Testing Procedure}

All benchmarks were run on a machine with two AMD Opteron processors with
2.20~GHz and 64~GB memory. All measurements reported in this paper
were obtained with \emph{SpiderMonkey JavaScript-C24.2.0}.

To evaluate our implementation we run the benchmark program in different
settings to separate the impact caused by the interpreter and the baseline
compiler. Recall that  no modifications were done to IonMonkey. Enabling IonMonkey in this state
would lead to meaningless results from executing a mixture of proxy-aware code and proxy-oblivious code. 
Therefore, \textbf{the IonMonkey compiler remains disabled}.

\begin{figure}[t]
  \centering
  \small
  \begin{tabular}{ l || I r | r || I r | r || I r | r H}
    \toprule
    \textbf{Benchmark}&
    \multicolumn{3}{c ||}{\textbf{Origin}}
    &
    \multicolumn{3}{c ||}{\textbf{Trap}}
    &
    \multicolumn{3}{c}{\textbf{Transparent}}
    \\
    &
    \textbf{Full}&
    \textbf{No-Ion}&
    \textbf{No-JIT}&
    \textbf{Full}&
    \textbf{No-Ion}&
    \textbf{No-JIT}&
    \textbf{Full}&
    \textbf{No-Ion}&
    \textbf{No-JIT}&
    \\
    \midrule
    Richards&
    11032
    &
    505
    &
    64.8
    &
    12854
    &
    502
    &
    63.3
    &
    10833
    &
    509
    &
    64.3
    &
    6361
    \\
    DeltaBlue&
    16344
    &
    453
    &
    82.5
    &
    16748
    &
    466
    &
    80.2
    &
    16436
    &
    466
    &
    79.6
    &
    6056
    \\
    Crypto&
    11430
    &
    817
    &
    111
    &
    11625
    &
    825
    &
    113
    &
    11485
    &
    793
    &
    109
    &
    11780
    \\
    RayTrace&
    33815
    &
    462
    &
    182
    &
    33484
    &
    455
    &
    173
    &
    34660
    &
    462
    &
    174
    &
    34796
    \\
    EarleyBoyer&
    16598
    &
    909
    &
    275
    &
    16267
    &
    938
    &
    271
    &
    16583
    &
    913
    &
    270
    &
    Unexpected token
    \\
    RegExp&
    1219
    &
    853
    &
    371
    &
    1239
    &
    842
    &
    362
    &
    1249
    &
    871
    &
    365
    &
    1267
    \\
    Splay&
    8858
    &
    802
    &
    409
    &
    8451
    &
    792
    &
    398
    &
    9693
    &
    857
    &
    409
    &
    No Termination
    \\
    SplayLatency&
    5279
    &
    1172
    &
    1336
    &
    5487
    &
    1222
    &
    1307
    &
    6518
    &
    1231
    &
    1338
    &
    No Termination
    \\
    NavierStokes&
    11639
    &
    841
    &
    155
    &
    11469
    &
    836
    &
    156
    &
    12714
    &
    834
    &
    148
    &
    12630
    \\
    pdf.js&
    6969
    &
    2759
    &
    704
    &
    6882
    &
    2764
    &
    697
    &
    6886
    &
    2793
    &
    691
    &
    RangeError
    \\
    Mandreel&
    12450
    &
    691
    &
    82.5
    &
    11977
    &
    711
    &
    82.4
    &
    12536
    &
    688
    &
    78.5
    &
    Out of Memory
    \\
    MandreelLatency&
    17131
    &
    3803
    &
    526
    &
    17014
    &
    3829
    &
    514
    &
    16785
    &
    3829
    &
    503
    &
    Out of Memory
    \\
    Gameboy Emulator&
    31203
    &
    4275
    &
    556
    &
    30928
    &
    4250
    &
    535
    &
    30339
    &
    4382
    &
    540
    &
    ReferenceError
    \\
    Code loading &
    8476
    &
    9063
    &
    9439
    &
    8457
    &
    9124
    &
    9318
    &
    8557
    &
    9114
    &
    9502
    &
    ReferenceError
    \\
    Box2DWeb&
    19197
    &
    1726
    &
    289
    &
    18941
    &
    1750
    &
    278
    &
    19051
    &
    1736
    &
    282
    &
    TypeError
    \\
    zlib&
    29208
    &
    28981
    &
    29052
    &
    29158
    &
    29097
    &
    29074
    &
    29041
    &
    28909
    &
    29108
    &
    ReferenceError
    \\
    TypeScript&
    19833
    &
    3708
    &
    1241
    &
    17474
    &
    3715
    &
    1210
    &
    19249
    &
    3666
    &
    1203
    &
    TypeError
    \\
    \midrule
    Total Score&
    12295
    &
    1594
    &
    456
    &
    12257
    &
    1604
    &
    447
    &
    12549
    &
    1610
    &
    445
    &
    \\
    \bottomrule
  \end{tabular}
  \caption{
    Scores for the Google Octane 2.0 Benchmark Suite (bigger is better).
    Column \textbf{Origin} gives the baseline scores for the unmodified engine.
    Column \textbf{Trap} shows the score values of the engine that implements the transparency trap and 
    column \textbf{Transparent} contains the scores for running the engine containing
    the transparent proxies.
    The sub-column \textbf{No-Ion} (\emph{no IonMonkey}) lists the scores with the baseline compiler enabled, 
    but with IonMonkey disabled. 
    Sub-column \textbf{No-JIT} (\emph{no just in-time compilation}) shows the scores of the interpreter
  without any kind of just in-time compilation.}
  \label{fig:performance}
\end{figure}

\subsubsection{Results}

Figure~\ref{fig:performance} contains the score values of all benchmark programs
in different configurations explained by the figure's caption. The examples show
the run-time impact of our modified engines vs.\ an unmodified engine. All
scores were taken from a deterministic run, which requires a predefined number
of iterations, and by using a warm-up run.

Comparing the total scores of the interpreter (column \textbf{No-JIT}), the
\textbf{Trap} version is 1.97\% slower than the unmodified engine and the
\textbf{Transparent} version is 2.41\% slower than the unmodified engine.
However, when comparing the total scores of the baseline compiler (column
\textbf{No-Ion}) we see that the \textbf{Trap} version is 0.62\% faster than the
unmodified engine and that the \textbf{Transparent} version is 1.00\% faster
than the unmodified engine.

At this point we have to mention that both differences are smaller than the
standard deviation of the mean total scores produced by an unmodified engine.
When measuring five runs with the same configuration we found a standard deviation
of 2.68 score points for the interpreter and a standard deviation of 23.44
points for the baseline compiler.

The numbers clearly show that both implementations do not have a statistically relevant impact on
the execution time of non-proxy code. This result is not surprising because the overwhelming
majority of equality comparisons have at least one non-object operand. As our modification only applies when
both operands are objects it is only exercised rarely (cf. Section~\ref{sec:implementation}) and
hence its performance impact is not measurable.

\subsection{An Analysis of Object Comparisons}
\label{sec:object_comparisons}

In this section, we answer \textbf{RQ2}, by considering how many object-object comparisons occur
during a normal program execution and how many of these may
fail when objects were wrapped by contracts implemented using opaque proxies.
To this end, we count object comparisons involving JavaScript Proxies
and give a classification that covers different types of object comparisons
whose result might be influenced by the transparency of the involved proxy
objects.



\subsubsection{The Testing Procedure}

For this experiment, we instrumented the JavaScript engine with a monitor to count and classify object-object
comparisons. Our subject programs are again taken from the Google Octane 2.0 Benchmark Suite.
Our wrapper model is a simple contract system which applies a dynamic type check (cf.\ Section~\ref{sec:contracts}) to the arguments of selected functions.

To prepare for the experiment,
a source-to-source translation generates for each function that occurs in
a program a new variant of the program, where exactly this
function is replaced by a function that wraps its arguments with a transparent
proxy. The proxy's handler implements a membrane. It
forwards the operation to the target and wraps its return value in another transparent
proxy. We rely on a weak map to avoid creating chains of nested proxies.

We applied this translation to the benchmark programs and executed each variant in our new engine, counting
and classifying each object comparison. As we used transparent proxies, normal execution of the
programs is not influenced. Because each function is wrapped individually, we can accurately detect
the effect of each single placement of a contract.



\subsubsection{Results}

First we introduce the types of object
comparisons that must be distinguished. We only consider comparisons between two objects where at
least one of them is a proxy object, because these are the only comparisons that may be affected if
the proxy is opaque. 

\begin{description}

  \item[\typeI]
    All comparisons between a proxy object and another object, which is either a
    native object or a proxy object from another membrane. Comparisons of this
    type always return \lstinline{false} when using opaque proxies. With transparent proxies the
    result is either \lstinline{true} or 
    \lstinline{false}, depending on the proxy's target object. 

  \item[\typeIa]
    The subset of \typeI, where the underlying target objects
    differ. Opaque and transparent proxies yield the same outcome,
    \lstinline{false}, but for different reasons.

  \item[\typeIb]
    The subset of \typeI, where the underlying target objects
    are the same. A comparison of this type yields \lstinline{false} when using
    opaque proxies, whereas transparent proxies yield \lstinline{true}.

  \item[\typeII]
    All comparisons between two proxy objects from the same membrane. If using an
    identity preserving membrane, a target object is only wrapped once. In this
    case the result will be \emph{true} if and only if they refer to the same target object,
    independent of the transparency of the proxies involved. Without 
    an identity preserving membrane, the use of opaque proxies yields
    \emph{false}, whereas the result with transparent proxies depends on the
    proxy's target object.

  \item[\typeIIa]
    The subset of \typeII, where the target objects differ. Opaque and
    transparent proxies yield the same outcome, \lstinline{false}, but for
    different reasons.

  \item[\typeIIb]
    The subset of \typeII, where both proxies refer to the same target object. A
    comparison of this type yields \lstinline{false} when using opaque proxies
    without an identity preserving membrane, whereas transparent proxies or
    an identity preserving membrane yield \lstinline{true}.

\end{description}


In this setting we count the comparison between two proxy objects from different 
membranes in category \typeI, because different contracts implement different membranes
and the mechanism that preserves the identity does not work when using different membranes.

Clearly, the {\typeIb} comparisons are the bad guys as they may flip. They are closely followed by
{\typeIIb} comparisons, although they are avoidable if identity preserving membranes are used throughout.

\begin{table}[t]
  \centering
  \small
  \begin{tabular}{ l || r | r | r | r | r}
    \toprule
    \textbf{Benchmark}&
    &
    \multicolumn{2}{c |}{\textbf{\typeI}}&
    \multicolumn{2}{c}{\textbf{\typeII}}\\
    &
    \textbf{Total}&
    \textbf{\typeIa}&
    \textbf{\typeIb}&
    \textbf{\typeIIa}&
    \textbf{\typeIIb}\\
    \midrule
    DeltaBlue&
    144126&
    29228&
    1411&
    33789&
    79698\\
    RayTrace&
    1075606&
    0&
    0&
    722703&
    352903\\
    EarleyBoyer&
    87211&
    8651&
    6303&
    53389&
    18868\\
    TypeScript&
    801436&
    599894&
    151297&
    20500&
    29745\\
    \bottomrule
  \end{tabular}
  \caption{Number of comparisons involving object proxies.
    Column \textbf{Total} contains the total number of comparisons.
    Column \textbf{Type-I} lists the comparisons of \text{\typeI}, divided in the two categories \text{\typeIa} and \text{\typeIb}.
    Column \textbf{Type-II} shows the number of \text{\typeII} comparisons, divided in the categories \text{\typeIIa} and \text{\typeIIb}.
  }
  \label{fig:comparisons}
\end{table}

Table~\ref{fig:comparisons} summarizes the number of comparisons
between native objects and proxy objects and among proxy objects. Comparisons between two primitive
values (e.g., a boolean, a number, a string, null, or  
undefined), comparisons between a primitive value and an object (proxy or native
object), and comparisons between two native objects are omitted from the results
because the result of the operation is not influenced by the
transparency of proxy objects.

Benchmark programs not listed in this table do not contain comparisons with a
proxy object: any function in the unlisted programs may be wrapped using any kind of membrane
without affecting its meaning. However, they still contain comparisons between native objects and
primitive values (usually the test \lstinline{ptr === null}).

The numbers in the table cover all comparison operators, namely \emph{equal}
(\lstinline{==}), \emph{not equal} (\lstinline{!=}), \emph{strict equal}
(\lstinline{===}), and \emph{strict not equal} (\lstinline{!==}). The meaning of
``the result is \emph{true}'' is generalized to the sense that \emph{equal} and
\emph{strict equal} will return \emph{true}, \emph{not equal} and \emph{strict
not equal} will return \emph{false}.

What we see in Table~\ref{fig:comparisons} is that there are three benchmarks with a non-negligible
number of bad {\typeIb} comparisons, although the majority of object comparisons is not
affected. The numbers also indicate that there are many more {\typeIIb} comparisons, so that any use
of non-identity preserving membranes should be strongly discouraged.

\subsection{Summary and Threats to Validity}

The evaluation shows that the implementation of a dynamic contract system based
on opaque proxies, whose monitoring replaces function arguments by proxy
objects, definitely influences the program execution (which answers \textbf{RQ2}), which in turn leads to
program errors.


The reason for the comparatively small number of flipped comparisons is due to the careful
handling of object comparisons in JavaScript. Results from previous unpublished experiments show
that approximately 6\% of all comparisons involve two objects. The vast majority of
comparisons either check an object against null or undefined, or compare
primitive values.


\section{Related Work}
\label{sec:related_work}


The JavaScript proxy API \cite{CutsemMiller2010,CutsemMiller2013} enables a
developer to enhance the functionality of objects easily. The implementation of
proxies opens up the means to fully interpose all operations on objects including
functions calls on function objects.

JavaScript proxies have been used for Disney's JavaScript contract system,
contracts.js \cite{Disney2013:contracts}, to enforce access permission contracts
\cite{KeilThiemann2013-Proxy}, as well as for other dynamic effects systems,
meta-level extension, behavioral reflection, security, or concurrency control
\cite{MillerCutsemTulloh2013,AustinDisneyFlanagan2011,BrachaUngar2004}.



Proxy-based implementations avoid the shortcomings of static implementations and
offline code transformations. In JavaScript, static approaches are often lacking
because of the dynamicity of the language. Proxies guarantee full interposition
and handle the full JavaScript language, including the
\lstinline{with}-statement, \lstinline{eval}, and arbitrary dynamic code loading
techniques.

The ideal contract system should not interfere with the normal execution of code as
long as the application code does not violate any contract. The application
should run as if no contracts were present \cite{DisneyFlanaganMcCarthy2011}.

Object equality becomes an issue for non-interference when contracts are implemented by some kind
of wrapper. The problem arises if an equality test between wrapper
and target or between different wrappers for the same target returns false
instead of true. This issue is known from other work involving wrappers for implementing object
extensions and multimethods \cite{WarthStanojevicMillstein2006,BaumgartnerJanscheLaufer2002}




The PLT group examines various designs for low-level mechanisms for implementing contracts and
related abstractions \cite{StricklandTobinHochstadtFindlerFlatt2012}. They propose two kinds of
proxies, chaperones and impersonators, that differ, for example, in
the degree of freedom for modifying the underlying object. They
experience similar problems with noninterference as we report in
Section~\ref{sec:use-case-contracts} when using the \lstinline{eq?} operator
which is similar to JavaScript's strict equality operator \lstinline{===} and roughly
implements pointer equality on objects. Racket's chaperones and impersonators are not transparent
with respect to this operator. However, the preferred equality operation in Racket, \lstinline{equal?},
implements structural equality which is indifferent to proxy transparency.
%
%
In contrast, JavaScript provides no built-in operation to test for structural equality, so that
developers need to build on pointer equality or roll their own structural equality.


\section{Conclusion}
\label{sec:conclusion}

Neither the transparent nor the opaque implementation of
proxies is appropriate for all use cases. We discuss several amendments and
propose two flexible solutions that enable applications requiring transparency as
well as opacity.
Both solutions are implemented as extensions of the SpiderMonkey JavaScript VM. This approach 
ensures full and transparent operation with all JavaScript programs. Hence, we can evaluate the
solution on real-world JavaScript programs.

A significant number of object comparisons would fail when mixing
opaque proxies and their target objects. This situation can arise when gradually adding contracts to
a program during debugging.
Identity preserving membranes decrease this number, but they are not able to
guarantee full noninterference.

We also measured the run-time impact of an implementation supporting transparent proxies on the
execution time of a realistic program mix. The results show that the modification to equality
required to support transparent proxies has no statistically significant impact on the execution time.



\appendix
\section{JavaScript's Equality Comparison}
\label{sec:algo-equality}

The interpreter's abstract equality comparison \lstinline{x==y}, called with
values x and y, produces \lstinline{true} or \lstinline{false}. The comparisons
works as follows~\cite{ecmascript2009}:
\begin{enumerate}
  \item If \lstinline{Type(x)} is the same as \lstinline{Type(y)}, then 
    \begin{enumerate}
      \item If \lstinline{Type(x)} is \lstinline{Undefined}, return \lstinline{true}.
      \item If \lstinline{Type(x)} is \lstinline{Null}, return \lstinline{true}.
      \item If \lstinline{Type(x)} is \lstinline{Number}, then 
        \begin{enumerate}
          \item If \lstinline{x} is \lstinline{NaN}, return \lstinline{false}.
          \item If \lstinline{y} is \lstinline{NaN}, return \lstinline{false}.
          \item If \lstinline{x} is the same Number value as \lstinline{y}, return \lstinline{true}.
          \item If \lstinline{x} is \lstinline{+0} and \lstinline{y} is \lstinline{-0}, return \lstinline{true}.
          \item If \lstinline{x} is \lstinline{-0} and \lstinline{y} is \lstinline{+0}, return \lstinline{true}.
          \item Return \lstinline{false}.
        \end{enumerate}
      \item If \lstinline{Type(x)} is String, then return true if \lstinline{x} and \lstinline{y} 
        are exactly the same sequence of characters (same length and same characters in corresponding 
        positions). Otherwise, return \lstinline{false}.
      \item If \lstinline{Type(x)} is Boolean, return \lstinline{true} if \lstinline{x} and \lstinline{y}
        are both \lstinline{true} or both \lstinline{false}. Otherwise, return \lstinline{false}.
      \textbf{\item If \lstinline{x} is Object, then
        \begin{enumerate}
          \item Let \lstinline{lhs} be the result of calling \lstinline{GetIdentityObject} on \lstinline{x}.
          \item Let \lstinline{rhs} be the result of calling \lstinline{GetIdentityObject} on \lstinline{y}.
          \item Return \lstinline{true} if \lstinline{lhs} and \lstinline{rhs} refer to the same object. 
            Otherwise, return \lstinline{false}.
        \end{enumerate}}
    \end{enumerate}
  \item If \lstinline{x} is \lstinline{null} and \lstinline{y} is \lstinline{undefined}, return \lstinline{true}.
  \item If \lstinline{x} is \lstinline{undefined} and \lstinline{y} is \lstinline{null}, return \lstinline{true}.
  \item If \lstinline{Type(x)} is Number and \lstinline{Type(y)} is String,
    return the result of the comparison \lstinline{x == ToNumber(y)}.
  \item If \lstinline{Type(x)} is String and \lstinline{Type(y)} is Number,
    return the result of the comparison \lstinline{ToNumber(x) == y}.
  \item If \lstinline{Type(x)} is Boolean, return the result of the comparison \lstinline{ToNumber(x) == y}.
  \item If \lstinline{Type(y)} is Boolean, return the result of the comparison \lstinline{x == ToNumber(y)}.
  \item If \lstinline{Type(x)} is either String or Number and \lstinline{Type(y)} is Object,
    return the result of the comparison \lstinline{x == ToPrimitive(y)}.
  \item If \lstinline{Type(x)} is Object and \lstinline{Type(y)} is either String or Number,
    return the result of the comparison \lstinline{ToPrimitive(x) == y}.
  \item Return \lstinline{false}.
\end{enumerate}
The algorithm starts with comparing values of the same type. In this branch it
first handles all comparisons of primitive values before it applies pointer
equality. After this point, the rest of the algorithm compares values of
different types.


\section{Getting the Identity Object}
\label{sec:algo-identity}


When the \lstinline{GetIdentityObject} internal method is called with argument
\lstinline{O} the following steps are taken:

\subsection{isTransparent-Trap}

\begin{enumerate}
  \item Assert either \lstinline{Type(O)} is Object and \lstinline{O} is not null.
  \item If \lstinline{O} is not a Proxy object, then return \lstinline{O}.
  \item Let \lstinline{handler} be the value of the [[ProxyHandler]] internal slot of \lstinline{O}.
  \item If \lstinline{handler} is \lstinline{null}, then throw a \lstinline{TypeError} exception.
  \item Let \lstinline{trap} be GetMethod(\lstinline{handler}, ``isTransparent'').
  \item ReturnIfAbrupt(\lstinline{trap}).
  \item Let \lstinline{trapResult} be the result of calling the [[Call]] internal method of \lstinline{trap} with \lstinline{handler} as the \lstinline{this} value.
  \item Let \lstinline{result} be ToBoolean(\lstinline{trapResult}).
  \item If \lstinline{result} is \lstinline{false} then return \lstinline{O}.
  \item Else, let \lstinline{target} be the value of the [[ProxyTarget]] internal slot of \lstinline{O}.
  \item Return the result of calling \lstinline{GetIdentityObject}, passing \lstinline{target} as the argument.
\end{enumerate}

\subsection{Transparent Proxy}

\begin{enumerate}
  \item Assert either \lstinline{Type(O)} is Object and \lstinline{O} is not null.
  \item If \lstinline{O} is not a TransparentProxy object, then return \lstinline{O}.
  \item Let \lstinline{target} be the value of the [[TransparentProxyTarget]] internal slot of \lstinline{O}.
  \item Return the result of calling \lstinline{GetIdentityObject}, passing \lstinline{target} as the argument.
\end{enumerate}

\section{Object.equals}
\label{sec:algo-equals}

\begin{enumerate}
  \item If number of arguments with which the function is called is less than 2, throw a \lstinline{TypeError} exception.
  \item If the number of arguments is 2 then:
    \begin{enumerate}
      \item Let \lstinline{lhs} be the result of calling \lstinline{GetIdentityObject} internal method on \lstinline{obj1}.
      \item Let \lstinline{rhs} be the result of calling \lstinline{GetIdentityObject} internal method on \lstinline{obj2}.
      \item Let \lstinline{result} be the value of testing the equality of the \lstinline{lhs} and \lstinline{rhs} references.
      \item Return the result of ToBoolean(\lstinline{result}).
    \end{enumerate}
  \item Else, let \lstinline{doRefEquality} be \lstinline{false}.
    \begin{enumerate}
      \item If \lstinline{obj1} is a transparent proxy, then let \lstinline{token} be the value of [[TokenObject]] internal slot of \lstinline{obj1}.
      \item If the reference of \lstinline{token} and \lstinline{secret} matches then let \lstinline{doRefEquality} be \lstinline{true}.
      \item If \lstinline{obj2} is a transparent proxy, then let \lstinline{token} be the value of [[TokenObject]] internal slot of \lstinline{obj2}.
      \item If the reference of \lstinline{token} and \lstinline{secret} matches then let \lstinline{doRefEquality} be \lstinline{true}.
      \item If \lstinline{doRefEquality} is true, let \lstinline{result} be the value of testing the equality of the \lstinline{obj1} and \lstinline{obj2} references.
      \item If \lstinline{doRefEquality} is \lstinline{false}:
        \begin{enumerate}
          \item Let \lstinline{lhs} be the result of calling \lstinline{GetIdentityObject} internal method on \lstinline{obj1}.
          \item Let \lstinline{rhs} be the result of calling \lstinline{GetIdentityObject} internal method on \lstinline{obj2}.
          \item Let \lstinline{result} be the value of testing the equality of the \lstinline{lhs} and \lstinline{rhs} references.
        \end{enumerate}
      \item Return the result of ToBoolean(\lstinline{result}).
    \end{enumerate}
\end{enumerate}


\bibliography{main}

\end{document}